\documentclass[fleqn,10pt]{wlscirep}
\usepackage[utf8]{inputenc}
\usepackage[T1]{fontenc}

\usepackage{cite}
\usepackage{amsmath,amssymb,amsfonts}
\usepackage{algorithmic}
\usepackage{graphicx}
\usepackage{textcomp}
\usepackage{multirow}
\usepackage{placeins}
\usepackage{tikz}
\usetikzlibrary{spy,calc}
\usepackage{verbatim}
\usepackage{booktabs,array} 
\usepackage{url}
\usepackage[ruled,vlined]{algorithm2e}
\usepackage{nicefrac}

\usepackage{hyperref}

\hypersetup{hidelinks,
backref=true,
pagebackref=true,
hyperindex=true,
breaklinks=true,
colorlinks=true,
linkcolor=blue,
urlcolor=blue,
bookmarks=true}

\usepackage{pgfplots}
\pgfplotsset{compat=newest, legend style={at={(1,0.05)},anchor=south east}}

\usepackage{cleveref}

\crefname{figure}{Figure}{Figure}
\crefname{table}{Table}{Table}
\crefname{equation}{equation}{Equation}

\usepackage{subcaption}
\usepackage{caption}
\newcommand{\beginsupplement}{%
        \setcounter{table}{0}
        \renewcommand{\thetable}{S\arabic{table}}%
        \setcounter{figure}{0}
        \renewcommand{\thefigure}{S\arabic{figure}}%
     }

\title{The effect of speech pathology on automatic speaker verification - a large-scale study\thanks{This study has been published in Scientific Reports, DOI: \url{https://doi.org/10.1038/s41598-023-47711-7}}}

\author[1,2,3,*]{Soroosh Tayebi Arasteh}
\author[1,2]{Tobias Weise}
\author[4]{Maria Schuster}
\author[1]{Elmar Noeth}
\author[1]{Andreas Maier}
\author[2]{Seung Hee Yang}
\affil[1]{Pattern Recognition Lab, Friedrich-Alexander-Universität Erlangen-Nürnberg, 91058 Erlangen, Germany}
\affil[2]{Speech \& Language Processing Lab, Friedrich-Alexander-Universität Erlangen-Nürnberg, 91054 Erlangen, Germany}
\affil[3]{Department of Diagnostic and Interventional Radiology, University Hospital RWTH Aachen, 52074 Aachen, Germany}
\affil[4]{Department of Otorhinolaryngology, Head and Neck Surgery, Ludwig-Maximilians-Universität München, 80333 Munich, Germany}

\affil[*]{soroosh.arasteh@fau.de}

\begin{abstract}
Navigating the challenges of data-driven speech processing, one of the primary hurdles is accessing reliable pathological speech data. While public datasets appear to offer solutions, they come with inherent risks of potential unintended exposure of patient health information via re-identification attacks. Using a comprehensive real-world pathological speech corpus, with over n$=$3,800 test subjects spanning various age groups and speech disorders, we employed a deep-learning-driven automatic speaker verification (ASV) approach. This resulted in a notable mean equal error rate (EER) of $0.89 \pm 0.06 \%$, outstripping traditional benchmarks. Our comprehensive assessments demonstrate that pathological speech overall faces heightened privacy breach risks compared to healthy speech. Specifically, adults with dysphonia are at heightened re-identification risks, whereas conditions like dysarthria yield results comparable to those of healthy speakers. Crucially, speech intelligibility does not influence the ASV system's performance metrics. In pediatric cases, particularly those with cleft lip and palate, the recording environment plays a decisive role in re-identification. Merging data across pathological types led to a marked EER decrease, suggesting the potential benefits of pathological diversity in ASV, accompanied by a logarithmic boost in ASV effectiveness. In essence, this research sheds light on the dynamics between pathological speech and speaker verification, emphasizing its crucial role in safeguarding patient confidentiality in our increasingly digitized healthcare era.

\end{abstract}

\begin{document}

\flushbottom
\maketitle

\thispagestyle{empty}


\section*{Introduction}

\subsection*{Background}

Speech is a biomarker that is extensively explored for the development of healthcare applications because of its low cost and non-invasiveness \cite{orozco2021there}. With the advances in deep learning (DL), data-driven methods have gained a lot of attention in speech processing in healthcare \cite{sztaho2019deep}.
For example, in the medical domain, speech biomarker reflects objective measurement that can be used for accurate and reproducible diagnosis. 
From diagnosis \cite{9053770, pappagari2020using, moro2018analysis, tayebiarasteh23_interspeech} to therapy\cite{jamal2017automatic, PoCaPCorpus_cu, harxiv.2001.04260}, pathological speech could be a rich source for different data-driven applications in healthcare. This is critical to the rapid and reliable development of medical screening, diagnostics, and therapeutics.
However, accessing pathological speech data for utilization in computer-assisted methods is a challenging and time-consuming process because of patient privacy concerns leading to the fact that most studies only investigated small cohorts due to the resulting lack of data \cite{maier2009speech}.

\subsection*{Related Works}

Pathological speech has garnered significant attention in DL-based automatic analyses of speech and voice disorders. Notably, Vásquez-Correa et al. \cite{8444654} broadly assessed Parkinson's disease, while Rios-Urrego et al. \cite{10.1007/978-3-031-40498-6_30} delved into evaluating the pronunciation skills of Parkinson's disease patients. Such works emphasize the potential of pathological speech as an invaluable resource for Parkinson's disease analysis. Additionally, numerous studies have employed pathological speech for DL-based analyses of Alzheimer’s disease. Pérez-Toro et al. \cite{10095219} illustrated the efficacy of the Arousal Valence plane for discerning and analyzing depression within Alzheimer’s disease. Pappagari et al. \cite{pappagari2020using} fused speaker recognition and language processing techniques for assessing the severity of Alzheimer's disease. Furthermore, García et al.'s work \cite{garcDysphonia} delved into dysphonia assessment, Kohlschein et al. \cite{8210766} addressed aphasia, Bhat et al. \cite{8962210} explored dysarthria, and Gargot et al. \cite{gargotautisms} investigated Autism Spectrum Disorders.

The burgeoning role of pathological speech in healthcare is evident, especially as computer-assisted, data-driven methods continue to flourish. However, this growth is tempered by the challenges in accessing pathological speech data. Patient privacy concerns make this not only a daunting task but also a protracted endeavor. Within this framework emerges a pivotal question: Does pathological speech, when examined as a biomarker, possess a heightened susceptibility to re-identification attacks compared to healthy speech? Addressing this necessitates the incorporation of ASV—a tool that verifies if an unrecognized voice belongs to a specific individual—to ascertain the privacy levels inherent to healthy speech data\cite{TOMASHENKO2022101362}. 

Laying the groundwork for understanding biomarkers in clinical research, Strimbu et al. \cite{strimbutravl} and Califf et al. \cite{califf2018biomarker} have proffered working definitions and established a foundational framework. Delving deeper, Marmar et al. \cite{speechusmarker} elucidated the diagnostic potential of speech-based markers, particularly in identifying posttraumatic stress disorder, while Ramanarayanan et al. \cite{ramanarayanan2022speech} unpacked both the opportunities and the impediments associated with harnessing speech as a clinical biomarker.
Remarkably, existing literature remains silent on the interplay between speech pathology and ASV. Our study is thus positioned to fill this void, venturing to discern the relative vulnerability of pathological speech to re-identification in contrast with its healthy counterpart.

\subsection*{Main Contributions}

In this study, we undertake a detailed look at how pathological speech affects ASV. We use a large and real-world dataset \cite{MAIER2009425} of around 200 hours of recordings that includes both pathological and healthy recordings. Our research focuses on text-independent speaker verification (TISV), to capture a broader range of scenarios~\cite{KINNUNEN201012, 10.1155/S1110865704310024}.  Considering the many factors that can sway ASV results, we made efforts to keep various conditions consistent by:

\begin{enumerate}
    \item Equalizing the training and test set sizes,
    \item ensuring consistent sound quality across recordings,
    \item matching age distributions within different subgroups,
    \item regulating background noise,
    \item controlling for the type of microphone utilized and the recording environment, and
    \item grouping by specific pathologies.
\end{enumerate}

In the sections that follow, we break down our findings methodically:

\begin{itemize}
    \item We start with broad-spectrum experiments to paint a comprehensive picture of our ASV system's prowess using the entire pathological dataset.
    \item Subsequently, our exploration narrows, dissecting the influence of specific pathologies on ASV for both adults and children.
    \item We then examine how combining data from different speech problems affects ASV. We also look into how the size of the training dataset influences ASV performance.
    \item Concluding our findings, we assess the influence of speech intelligibility on ASV's performance.
\end{itemize}

We assume that equal error rate (EER) is a measure of anonymity in the dataset. The lower the EER, the higher the vulnerability of the respective group. This is also a common choice in speaker verification challenges \cite{TOMASHENKO2022101362}. Furthermore, we use word recognition rate (WRR) as a measure of speech intelligibility as it demonstrated high and significant correlations in many previous studies \cite{MAIER2009425, kitzing2009automatic,maierclpfuly,maier2009speech}. The lower the WRR the less intelligible, the speech of the persons in the respective group. 

Our goal is to uncover the connection between pathological speech conditions and speaker verification’s success rate. We show evidence that the distinct features of pathological speech, when paired with different recording conditions, influence speaker verification outcomes.


\begin{table}[t]
\centering

\begin{tabular}{>{\arraybackslash}m{2.25cm}>{\centering\arraybackslash}m{1.45cm}>{\centering\arraybackslash}m{1.75cm}>{\centering\arraybackslash}m{1.5cm}>{\centering\arraybackslash}m{1.6cm}>{\centering\arraybackslash}m{2.05cm}>{\centering\arraybackslash}m{2.2cm}>{\centering\arraybackslash}m{2.09cm}}
\toprule
\textbf{} & \textbf{Total num speakers} & \textbf{Num female speakers} & \textbf{Num male speakers} & \textbf{Total num utterances} & \textbf{Total duration [hours]} & \textbf{Mean $\pm$ std dev age [years]} & \textbf{Mean $\pm$ std dev WRR [\%]} \\
\midrule
\textbf{Adults} & &  &    &  &  &  \\
dysglossia-dnt & $883$ & $245$ & $638$  & $21,338$ & $41.21$ & $60.91\pm 11.95$ & $62.61\pm 15.96$\\
dysarthria-plant & $533$ & $258$ & $275$  & $7,128$ & $13.37$ & $62.70\pm 15.29$ & $69.11\pm 12.70$ \\
dysphonia-logi & $86$ & $10$ &  $76$ & $900$ & $1.67$ & $59.28\pm 10.67$ & $51.78\pm 15.84$\\
\textit{sum patients} & \textit{1,502} & \textit{513} &  \textit{989} & \textit{29,366} & \textit{56.25} & $\textit{61.40}\pm \textit{13.34}$ & $\textit{63.37}\pm \textit{15.78}$\\
\midrule
ctrl-plant-A & $85$ & $42$ & $43$ & $891$ & $1.60$ & $23.93\pm 15.62$ & $73.72\pm 15.69$\\
\midrule
\midrule
\textbf{Children} & &  &    &  &  &  \\
CLP-dnt & $476$ & $216$ & $260$  & $16,964$ & $34.63$ & $9.69\pm 3.98$ & $48.28\pm 17.30$\\
CLP-plant & $124$ & $58$ & $66$  & $4,120$ & $7.88$ & $9.27\pm 2.58$  & $57.61\pm 13.86$\\
\textit{sum patients} & \textit{600} & \textit{264} & \textit{326}  & \textit{21,084} & \textit{42.51} & $\textit{9.58} \pm \textit{3.71}$ & $\textit{50.12} \pm \textit{17.07}$ \\
\midrule
ctrl-plant-C & $1,662$ & $900$ &  $761$ & $54,896$ & $98.46$ & $12.16\pm 3.72$ & $65.87\pm 12.44$ \\
\bottomrule
\end{tabular}
\caption{Dataset statistics used in this study. The table provides details on the total number of speakers, gender distribution, utterance count, total duration in hours, age range, and word recognition rates (WRRs). The corpus is divided into two groups: adults (those aged over 20 years) and children (those aged 20 years or younger). Both groups encompass control subsets ("ctrl-plant") comprising healthy subjects. Abbreviations are as follows: dysglossia: Patients with dysglossia who had prior maxillofacial surgery before assessment; dysarthria: Patients diagnosed with dysarthria; dysphonia: Patients with voice disorders; CLP: Children diagnosed with cleft lip and palate; dnt: Recordings using the "dnt Call 4U Comfort" headset \cite{maierclpfuly}; plant: Recordings using a specific Plantronics Inc. headset \cite{plantronicscite}; logi: Recordings using a specific Logitech International S.A. headset \cite{logitechcite}; ctrl: Control group. The suffix "-A" denotes the adult subset, whereas "-C" pertains to the children subset. Age and WRR values are expressed as mean $\pm$ standard deviation.}
\label{tab:nordwindplakssstat}
\end{table}

\begin{table}[t]
\centering

\begin{tabular}{>{\arraybackslash}m{2.7cm}>{\centering\arraybackslash}m{2.5cm}>{\centering\arraybackslash}m{2.6cm}>{\centering\arraybackslash}m{4.1cm}>{\centering\arraybackslash}m{3.6cm}}
\toprule
\textbf{Experiment name} & \textbf{Total num speakers} & \textbf{Subset} & \textbf{Pathology} & \textbf{Microphone}  \\
\midrule
dysglossia-dnt-85 & $85$ & Adults & Dysglossia  & dnt Call 4U \\
dysarthria-plant-85 & $85$ & Adults & Dysarthria  & Plantronics \\
dysphonia-logi-85 & $85$ & Adults &  Dysphonia & Logitech \\
ctrl-plant-A-85 & $85$ & Adults & None (healthy) & Plantronics \\
CLP-dnt-124 & $124$ & Children & CLP  & dnt Call 4U \\
CLP-plant-124 & $124$ & Children & CLP  & Plantronics \\
ctrl-plant-C-124 & $124$ & Children &  None (healthy) & Plantronics \\
all-children-124 & $124$ & Children &  Mixture of CLP \&  healthy & dnt Call 4U \& Plantronics \\
CLP-dnt-plant-500 & $500$ & Children & CLP  & dnt Call 4U \\
ctrl-plant-C-500 & $500$ & Children &  None (healthy) & Plantronics \\
all-spk-50 & $50$ & Adults \& Children & Mixture of all \& healthy  & Mixture of all \\
all-spk-500 & $500$ & Adults \& Children & Mixture of all \& healthy  & Mixture of all \\
all-spk-1500 & $1,500$ & Adults \& Children &  Mixture of all \& healthy & Mixture of all  \\
all-spk-3000 & $3,000$ & Adults \& Children & Mixture of all \& healthy & Mixture of all \\

\bottomrule
\end{tabular}
\caption{Overview of the experiments performed in this study. Abbreviations: dysglossia: Patients with dysglossia who underwent prior maxillofacial surgery; dysarthria: Patients diagnosed with dysarthria; dysphonia: Patients with voice disorders; CLP: Children with cleft lip and palate; dnt: Recordings from the "dnt Call 4U Comfort" headset \cite{maierclpfuly}; plant: Recordings via Plantronics Inc. headset \cite{plantronicscite}; logi: Recordings via Logitech International S.A. headset \cite{logitechcite}; ctrl: Control group. The labels "-A" and "-C" respectively indicate adult and children subsets. Numbers appended, such as "-85" in "dysglossia-dnt-85", represent the total speaker count for that experiment. "all-spk" designates experiments combining all dataset speech signals from both adults and children, and both pathological and healthy subjects.}
\label{tab:expalantion}
\end{table}


\section*{Results}

\subsection*{Pathology Influences ASV Performance}

When examining pathological recordings from both adult and child subsets, our results showed a mean EER of $0.89 \pm 0.06$\%. For this, n$=$2,064 speakers were used for training and n$=$517 for testing. Notably, this EER is lower than common values found in datasets such as LibriSpeech \cite{7178964} or VoxCeleb1\&2 \cite{Nagrani17, Chung18b}. This outcome was the average from 20 repeated experiments to counteract the potential biases of random sampling. 
To ensure an equitable comparison across groups, each subgroup was adjusted in terms of age distribution and speaker numbers. After employing standard training and evaluation, we then evaluated the speaker verification outcomes for each subgroup against control groups.

\begin{figure}[t]
    \centering
        \includegraphics[scale=0.78]{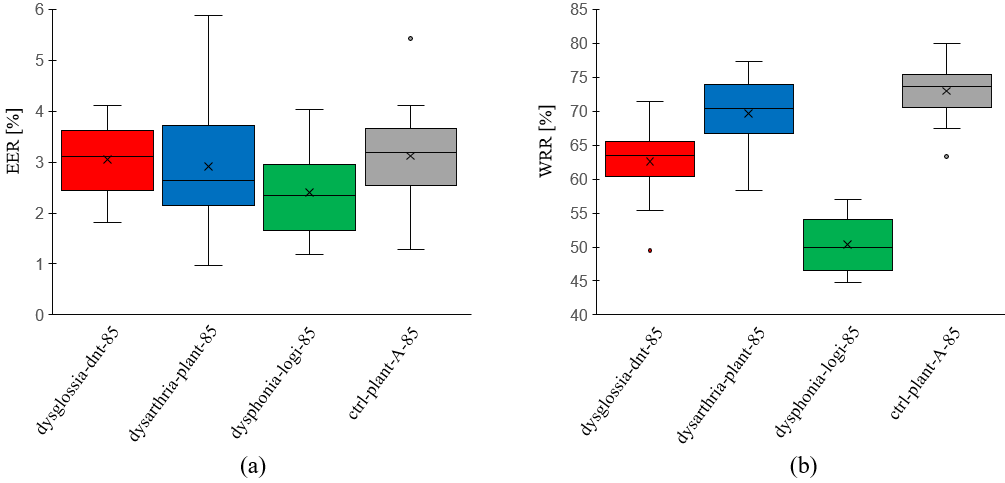}
    \caption{Evaluation results of speaker verification on the adults for individual groups for 20 repetitions. During each repetition, n$=$85 speakers are sampled for each group and n$=$68 of them were assigned to training and n$=$17 speakers to test. All the values are given in percent. (a) Equal error rate (EER) values. (b) Word recognition rate (WRR) values. Abbreviations: dysglossia: Patients with dysglossia who underwent prior maxillofacial surgery; dysarthria: Patients diagnosed with dysarthria; dysphonia: Patients with voice disorders; CLP: Children with cleft lip and palate; dnt: Recordings from the "dnt Call 4U Comfort" headset \cite{maierclpfuly}; plant: Recordings via Plantronics Inc. headset \cite{plantronicscite}; logi: Recordings via Logitech International S.A. headset \cite{logitechcite}; ctrl: Control group. Numbers appended, such as "-85" in "dysglossia-dnt-85", represent the total speaker count for that experiment.}
    \label{fig:nordwind_results}
\end{figure}

Adults: Adult patients were divided into three categories: "dysglossia-dnt", "dysarthria-plant", and "dysphonia-logi". For benchmarking, $n=85$ healthy individuals formed the control group, labeled as "ctrl-plant-A". 
When examining EER values, both "dysglossia-dnt-85" ($3.05 \pm 0.74\%$) and "dysarthria-plant-85" ($2.91 \pm 1.09\%$) showed no significant difference from the control group "ctrl-plant-A-85" ($3.12 \pm 0.94\%$) with P$= 0.786$ and P$=0.520$, respectively. In contrast, the "dysphonia-logi-85" group, at an EER of $2.40 \pm 0.84\%$, was significantly different from the control, with a P$= 0.015$. Refer to \cref{fig:nordwind_results} for a visual representation of these findings.

\begin{figure}[t]
    \centering
        \includegraphics[scale=0.4]{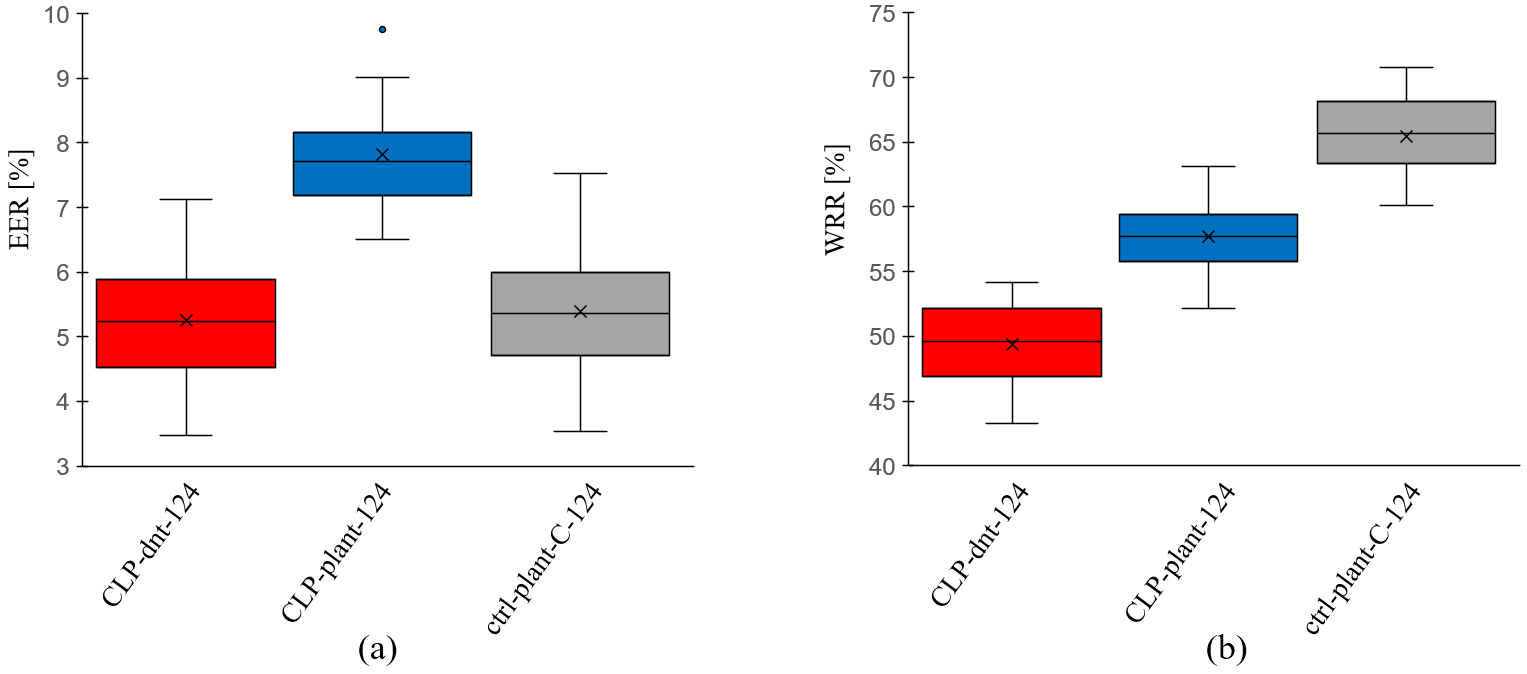}
    \caption{Evaluation results of speaker verification on the children for individual groups for 20 repetitions. During each repetition, 124 speakers are sampled for each group and 99 of them were assigned to training and 25 speakers to test. All the values are given in percent. (a) Equal error rate (EER) values. (b) Word recognition rate (WRR) values. CLP: Children with cleft lip and palate; dnt: Recordings from the "dnt Call 4U Comfort" headset \cite{maierclpfuly}; plant: Recordings via Plantronics Inc. headset \cite{plantronicscite}; ctrl: Control group. Numbers appended, such as "-124" in "CLP-dnt-124", represent the total speaker count for that experiment.}
    \label{fig:plakss_results}
\end{figure}

Children: Children were divided into two categories, "CLP-dnt" and "CLP-plant", both representing patients with cleft lip and palate (CLP). Additionally, a control group of $n=1,662$ healthy children, "ctrl-plant-C", was used. As for the EER values, "CLP-dnt-124" yielded ($5.25 \pm 0.90\%$, which was not significantly different from the control group's $5.72 \pm 1.05$ (P$= 0.134$). However, "CLP-plant-124" stood out at $7.82 \pm 0.91\%$, differing significantly from the control's $5.72 \pm 1.05\%$ (P$< 0.001$). \cref{fig:plakss_results} offers a detailed visual comparison among the groups.

\subsection*{Pathological Diversity in Speakers Leads to Substantial Reduction in ASV Error Rate}

In our pursuit to understand the influence of pathological diversity on ASV, various datasets were combined, maintaining the speaker count for training and testing as for the children (see \cref{tab:expalantion}), with an age distribution to match. Upon combining the variations from the "all-children-124" set, we noticed a notable improvement in average EER. Specifically, it stood at $4.80 \pm 0.98\%$, which was considerably better than the control group "ctrl-plant-C-124" that recorded an EER of $5.72 \pm 1.05\%$ (P$= 0.006$). This data highlights the potential benefits of integrating multiple sources of variation in reducing error rates. 

Further, when leveraging larger training sets infused with pathological diversity (see \cref{tab:expalantion}), the EER for the mixed pathological group "CLP-dnt-plant-500" was $2.88 \pm 0.25\%$. In comparison, the EER for the healthy group "ctrl-plant-C-500" was $3.04 \pm 0.17\%$ (P$= 0.020$). This reinforces the premise that the pathological group, with its inherent diversity, offers an advantage in speaker verification over the relatively homogenous healthy group.

\subsection*{Increase in Training Speaker Number Yields Logarithmic Enhancement in ASV Performance}

Exploring the impact of training set size on ASV performance, we integrated both pathological and healthy speakers from a comprehensive pool of n$=$3,849. Various speaker groups were drawn from this collective, and they underwent our standard training and evaluation processes.

The "all-spk-50" dataset, which comprised 50 speakers, recorded an EER of $5.19\pm 1.63\%$. With an increased speaker count in the "all-spk-500" dataset, the EER was reduced to $1.87\pm 0.19\%$, marking a significant improvement with a P$< 0.001$. Extending the dataset to 1,500 speakers ("all-spk-1500"), the EER further decreased to$1.15\pm 0.10\%$, surpassing the performance of the previous group with a P$< 0.001$. When the dataset was expanded to 3,000 speakers ("all-spk-3000"), the EER diminished to $0.90\pm 0.05\%$, outperforming the 1,500-speaker dataset with a P$< 0.001$.

\begin{figure}[t]
    \centering
        \includegraphics[scale=0.355]{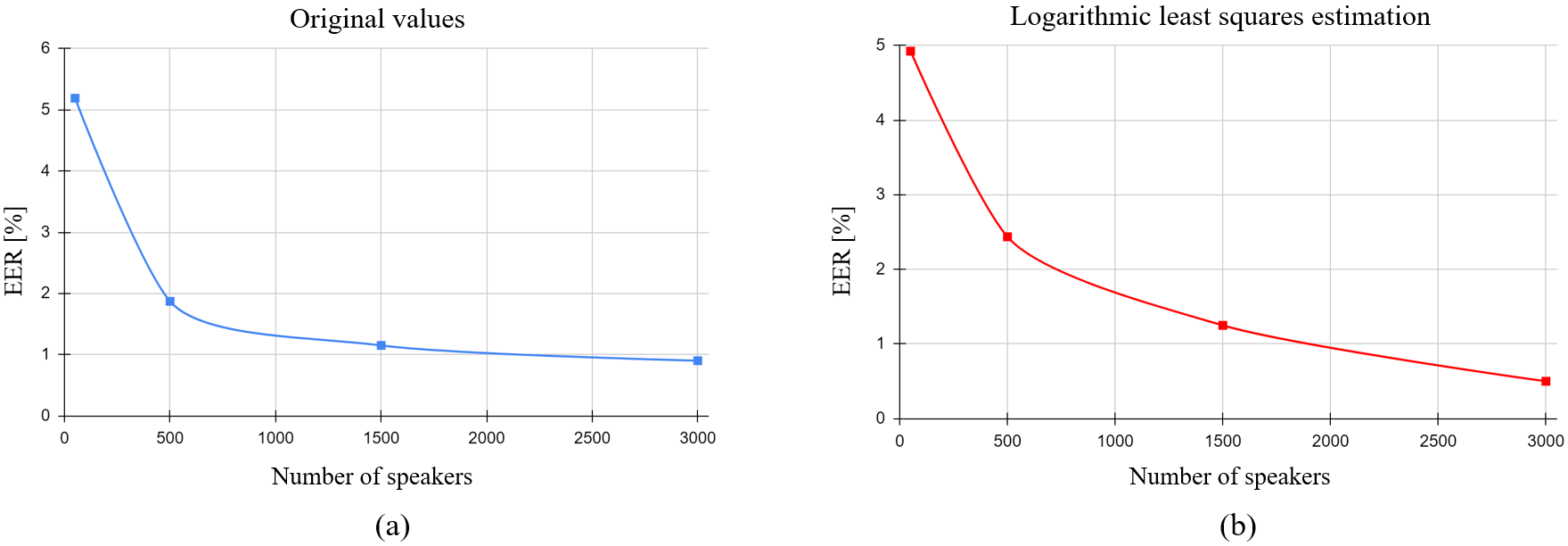}
    \caption{EER results utilizing different training speaker numbers. (a) The original values. The EER values are 5.19, 1.87, 1.15, and 0.90 for the cases with n$=$50, 500, 1,500, and 3,000 speakers, respectively. (b) The resulting curve after logarithmic least squares regression according to $y=9.1543237903 - 1.0809973418 \cdot \ln{x}$. The regression coefficient of determination (R$^2$) equals 0.95. We observe that increasing total training speaker numbers, leads to logarithmic improvement of the ASV performance.}
    \label{fig:logarithm_effect}
\end{figure}

This decrement in EER as the number of training speakers increased is visually captured in \cref{fig:logarithm_effect}, which underscores the logarithmic reduction of the error rate with an augmented training set size.

\subsection*{Intelligibility of Patients Is not an Influencing Factor in ASV}

\begin{figure}[t]
    \centering
        \includegraphics[scale=1.0]{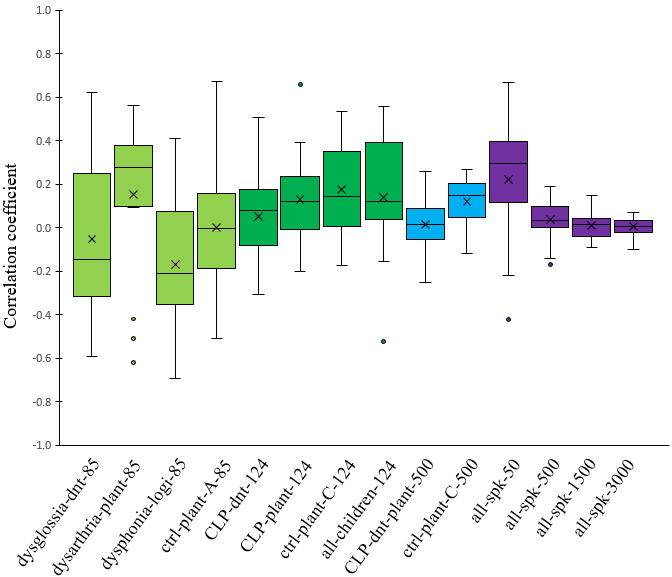}
    \caption{Correlation coefficients between EER values and WRR values for all the experiments. Abbreviations: dysglossia: Patients with dysglossia who underwent prior maxillofacial surgery; dysarthria: Patients diagnosed with dysarthria; dysphonia: Patients with voice disorders; CLP: Children with cleft lip and palate; dnt: Recordings from the "dnt Call 4U Comfort" headset \cite{maierclpfuly}; plant: Recordings via Plantronics Inc. headset \cite{plantronicscite}; logi: Recordings via Logitech International S.A. headset \cite{logitechcite}; ctrl: Control group. The labels "-A" and "-C" respectively indicate adult and children subsets. Numbers appended, such as "-85" in "dysglossia-dnt-85", represent the total speaker count for that experiment. "all-spk" designates experiments combining all dataset speech signals from both adults and children, and both pathological and healthy subjects.}
    \label{fig:correlation_effect}
\end{figure}

To explore the relationship between intelligibility of the speakers and ASV, we computed correlation coefficients between EER results (representing speaker verification metric) and WRR values (indicating speech intelligibility) of all the experiments. \cref{fig:correlation_effect} illustrates the correlation coefficients between error rates and recognition rates of all the experiments. We observed that the correlation coefficients in all cases were very small. Notably, as the number of speakers increased, this correlation diminished even further. Specifically, in the "all-spk-50" experiment—wherein all healthy and pathological speech signals from both children and adults were fused and a random sample of 50 speakers was taken—the correlation coefficient between EER and WRR stood at $0.22\pm 0.30$. For larger sample sizes, "all-spk-500" had a coefficient of $0.04\pm 0.09$, "all-spk-1500" showed $0.01\pm 0.06$, and the largest sample, "all-spk-3000", exhibited an almost non-existent correlation of $0.00\pm 0.04$. This data strongly indicates that the intelligibility of a patient's speech does not wield substantial influence over the performance of an ASV system.

\section*{Discussion}

This study, drawing from in-depth analysis of recordings of both pathological and healthy subjects, offers strong evidence that certain speech pathologies might serve as viable biomarkers in automatic speaker verification (ASV). Intriguingly, certain pathological speech forms demonstrated a heightened vulnerability, shedding light on the potential risks associated with patient re-identification. Using a state-of-the-art deep learning framework for training and evaluation, our research dove deep into these complexities.

To objectively gauge the impact of pathology on ASV, rigorous controls were established to address potential confounders, such as age distribution, recording conditions, microphone types, audio clarity, and speech intelligibility. Analyzing pathological recordings from n$=$2,581 adults and children, the results illustrated a mean EER of $0.89 \pm 0.06$\%. Strikingly, this EER is appreciably lower than that in non-pathological datasets like LibriSpeech \cite{7178964} or VoxCeleb1\&2 \cite{Nagrani17, Chung18b}. To circumvent biases from random sampling, we derived this result from an average of 20 repeated trials.

Data from children yielded intriguing insights. Pathological children, on average, exhibited higher EER values than their healthy peers. For instance, the "CLP-plant-124" subgroup displayed a $27\%$ surge in EER, under identical recording conditions as the control group. Conversely, adult data showed decreased error rates for those with speech pathologies. This disparity could stem from the ASV model's inclination towards adult speech patterns, coupled with the evolving nature of children's speech influenced by cognitive development.

Our exploration of the relationship between speech pathologies and ASV efficacy yielded further illuminating findings. The integration of diverse pathological voices into the dataset notably enhanced ASV accuracy. For example, the average EER experienced a significant improvement when varied pathologies from the "all-children-124" set were included, performing better than the control group. This suggests that incorporating multiple sources of variability could be pivotal in refining ASV outcomes.

Moreover, the trend of enhanced ASV performance persisted when the training sets were enriched with pathological diversity. For instance, the mixed pathological group's EER was lower than that of the healthy group, emphasizing the potential advantage of pathological diversity in speaker verification.

Delving into the effects of training set size on ASV, we observed that expanding the speaker pool, to include both pathological and healthy voices, consistently boosted ASV accuracy. For example, with the increase in speakers in datasets like "all-spk-500", "all-spk-1500", and "all-spk-3000", there was a consistent drop in EER. Such an incremental improvement with increasing dataset size hints at the potential of large datasets in drastically enhancing ASV efficacy.

Diving deeper into the potential variables that could influence ASV, we probed the intricacies of speech intelligibility. Analyzing the correlation between EER results (indicating ASV performance) and WRR values (indicating speech intelligibility) across experiments, we uncovered intriguing patterns. The consistently minimal correlation values, especially in larger speaker samples, unequivocally underline that a speaker's intelligibility does not significantly sway ASV system outcomes. This observation challenges the often-presumed importance of speech clarity in ASV systems, suggesting that even if a speaker's utterances are not distinctly clear, it might not substantially hamper the system's verification accuracy. This revelation could have profound implications, especially in scenarios where speech anomalies are prevalent.

Our study stands out due to its novel emphasis on the intersection of speech pathologies and ASV. While a significant portion of recent ASV research has dedicated efforts to improve algorithms and tackle speaker verification challenges by utilizing well-established non-pathological datasets—such as LibriSpeech \cite{7178964} (EER: $3.85\%$ on the 'test-clean' subset with n$=$40 test speakers and $3.66\%$ on 'test-other' subset with n$=$33 test speakers \cite{arasteh2020generalized}), VoxCeleb 1 \cite{Nagrani17} (EER: $7.80\%$ with n=40 test speakers), and VoxCeleb 2 \cite{Chung18b} (EER: $3.95\%$ with n$=$40 test speakers)—there is a conspicuous absence of studies that delve into the relationship between speech pathologies and ASV. In our initial exploration, we identified a substantially low mean EER of $0.89 \pm 0.06$\% when analyzing pathological speech patterns. While our research introduces a unique dimension to ASV by examining speech pathologies, our results are not directly comparable to those derived from non-pathological conventional datasets because of the inherent differences in the characteristics and challenges posed by pathological speech patterns, recording conditions, testing criteria, text-independent or dependent nature of ASV task, etc. Nonetheless, our study lays the groundwork for a more profound understanding of ASV systems, particularly in contexts permeated by speech anomalies.

Our study had limitations. First, due to the constrained availability of adult subjects, we were unable to harmonize age distributions among individual adult sub-groups, potentially narrowing the generalizability of our findings within adult demographics. To enhance clarity and depth in comparative results, securing additional utterances from both patient and healthy adult populations in future studies is paramount. Second, despite utilizing a robust, large-scale dataset sourced from an extensive array of participants, our pathological corpus \cite{MAIER2009425} was circumscribed to specific speech pathologies and voice disorders, namely dysglossia following maxillofacial surgery, dysarthria, dysphonia, and cleft lip and palate. Subsequent research could potentially broaden this dataset to encompass additional conditions such as aphasia \cite{8210766}. Furthermore, our pathological corpus \cite{MAIER2009425}, though diverse in its recording locations – spanning cities like (i) Erlangen, Bavaria, Germany, (ii) Nuremberg, Bavaria, Germany, (iii) Munich, Bavaria, Germany, (iv) Stuttgart, Baden-Württemberg, Germany, and (v) Siegen, North Rhine-Westphalia, Germany – exclusively features German-language utterances. While we expect that language may not correlate with the susceptibility of pathological speech to re-identification, it remains essential to confirm these findings across multiple languages to validate and generalize our results. Lastly, although we have illuminated the effects of speech pathology across distinct pathology and voice disorder groupings, an important area warranting deeper exploration is the examination at an individual level. In our future direction, this will be a focal area of emphasis.

In conclusion, our findings elucidate the complex relationship between specific speech pathologies and their impact on ASV. We have pinpointed pathologies such as dysphonia and CLP as warranting increased attention due to their amplified re-identification risks. Contrary to prevalent beliefs, our study also reveals that pristine speech clarity is not pivotal for ASV's effective operation. The diversity of datasets plays a crucial role in augmenting ASV performance, a noteworthy insight for future ASV developments. However, as the demand for open-source speech data rises, our study emphasizes the critical need for the development or refinement of anonymization techniques. While research in the domain of anonymization is evolving, as indicated by works like \cite{TOMASHENKO2022101362, perero2022x, yoo2020speaker, srivastava2020design}, there remains a pressing need for techniques specifically attuned to pathological speech. It is imperative for the scientific community to strike a harmonious balance between maximizing the utility of data and safeguarding the privacy and rights of individuals.


\section*{Methods}

\subsection*{Ethics Declarations}

The study and the methods were performed in accordance with relevant guidelines and regulations and approved by the University Hospital Erlangen’s institutional review board (IRB) with application number 3473. Informed consent was obtained from all adult participants as well as from parents or legal guardians of the children.

\subsection*{Pathological Speech Corpus}

Initially, we gathered a total of $216.88$ hours of recordings from n$=$4,121 subjects using PEAKS \cite{MAIER2009425}, a prominent open-source tool.
Given PEAKS' extensive use in scientific circles across German-speaking regions since 2009, its database offers a comprehensive assortment of recordings reflecting a multitude of conditions. To arrive at the finalized dataset, the following steps of intricate analysis were executed:

(i) Recordings missing data points such as WRR, diagnosis, age, microphone, or recording environment were purged from the collection. (ii) Recordings that were noisy or of poor quality were also discarded. (iii) Any data categorized as 'test' or deemed irrelevant by examiners were omitted. (iv) Segments of recordings containing the examiner's voice or those from multiple speakers were excised. (v) Leveraging PEAKS' ability to automatically segment recordings into shorter utterances (ranging from 2 to 10 seconds based on voice activity), speakers that, post these steps, were left with fewer than 8 utterances were excluded. (vi) Finally, recognizing age as a potentially influential variable, the dataset was bifurcated into two major categories: adults and children. This segregation was vital to ensure nuanced analyses given the distinctive characteristics and potential performance deviations associated with these age groups.

In the end, a total of n$=$3,849 participants were included in this study.
\cref{tab:nordwindplakssstat} shows an overview and the statistics of the data subsets, i.e., the adults and children. The utilized dataset contained $198.82$ hours of recordings from n$=2,102$ individuals with various pathologies and n$=$1,747 healthy subjects. To ensure our results are reliable, we carefully sorted these recordings based on pathology types and recording settings.
The utterances were recorded at $16 \text{\,kHz}$ sampling frequency and 16 bit resolution \cite{MAIER2009425}. All the subjects were native German speakers, using different dialects including the standard German ("Hochdeutsch") as well as local dialects.

\subsubsection*{Adults}
Subjects above the age of 20 were included in the adults subset of our dataset. n$=$1,502 patients read "Der Nordwind und die Sonne", the German version of the text "The North Wind and the Sun", a fable from Aesop. It is a phonetically rich text with 108 words, of which $71$ are unique \cite{MAIER2009425}. Our adult patient cohort had an age range of $21$ to $94$ years (mean $61.40 \pm 13.34$ and median $62.49$). \cref{fig:nordwind_hist}a shows the age histogram of the three patient groups of adults used in this study ("dysglossia-dnt", "dysarthria-plant", and "dysphonia-logi").

"dysglossia-dnt" represents the group of patients who had dysglossia, underwent a maxillofacial surgery before the pathology assessment, and all were recorded using the "dnt Call 4U Comfort" headset \cite{maierclpfuly}. Out of all the available utterances, we selected those that were recorded using the same microphone. "dysarthria-plant" is a group of patients who had dysarthria and underwent speech therapy and all were recorded using a specific headset from Plantronics Inc.\cite{plantronicscite}. "dysphonia-logi" represents the patients who had voice disorders and all were recorded using a specific headset from Logitech International S.A.\cite{logitechcite}. Finally, as a control group ("ctrl-plant-A"), n$=$85 healthy individuals were asked to undergo the test using the same Plantronics headset\cite{plantronicscite}.

\subsubsection*{Children}

\begin{figure}[t]
    \centering
        \includegraphics[scale=0.74]{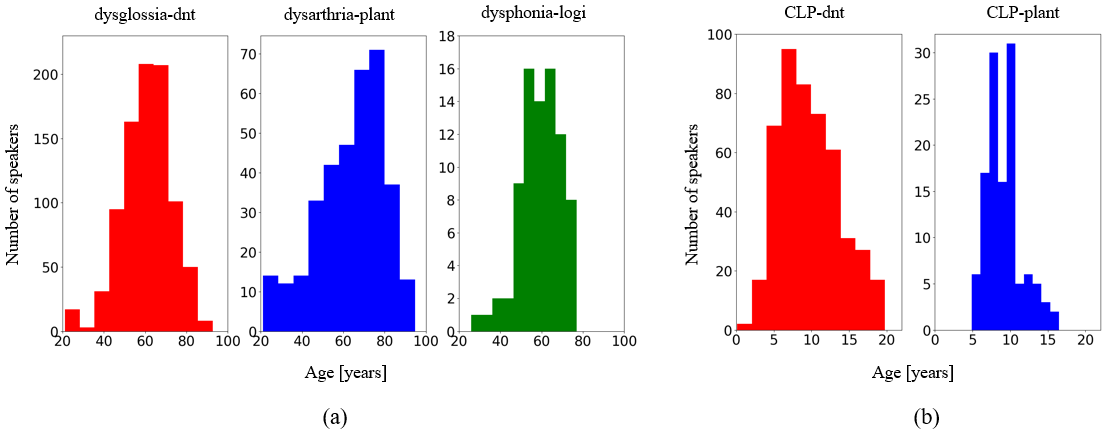}
    \caption{Age histograms of the patient groups. (a) The adults; (b) The children group. Abbreviations: dysglossia: Patients with dysglossia who underwent prior maxillofacial surgery; dysarthria: Patients diagnosed with dysarthria; dysphonia: Patients with voice disorders; CLP: Children with cleft lip and palate; dnt: Recordings from the "dnt Call 4U Comfort" headset \cite{maierclpfuly}; plant: Recordings via Plantronics Inc. headset \cite{plantronicscite}; logi: Recordings via Logitech International S.A. headset \cite{logitechcite}.}
    \label{fig:nordwind_hist}
\end{figure}

Six hundred children with an age range of $2 - 20$ years old (mean $9.58 \pm 3.71$ and median $9.12$) were included in the study. The test consisted of slides that showed pictograms of the words to be named. In total, the test contained 97 words which included all German phonemes in diﬀerent positions. Due to the fact that some children tended to explain the pictograms with multiple words, and some additional words were uttered in between the target words, the recordings were automatically segmented at pauses that were longer than $1$s \cite{MAIER2009425}. \cref{fig:nordwind_hist}b illustrates the age histogram of the two patient groups of children used in this study ("CLP-dnt" and "CLP-plant").

"CLP-dnt" represents children with cleft lip and palate (CLP), which is the most common malformation of the head with incomplete closure of the cranial vocal tract \cite{Wantia2002TheCU, Millardclefd, harding1996characteristics, maierclpfuly}, which all were recorded using the same "dnt Call 4U Comfort" headset \cite{maierclpfuly} as for the adults.
Finally, as a control group ("ctrl-plant-C"), n$=$1,662 healthy children were asked to undergo the test with similar recording conditions as in "ctrl-plant-A".

\subsection*{Experimental Design}

\cref{tab:expalantion} shows an overview of the different experiments performed in this study.

\subsubsection*{Analysis of Impact of Pathology on ASV Performance}

Initially, the study aimed to analyze the performance of automatic speaker verification (ASV) systems on recordings from individuals with various speech pathologies. For each category of adults, recordings were sourced from 85 predetermined speakers. As reflected in \cref{tab:nordwindplakssstat}, a precise age match for adults was challenging due to the limited recordings available. Nonetheless, $20\%$ of the speakers were assigned to the test set and $80\%$ to the training set. This selection and allocation process was iterated $20$ times. For the children's group, given the limited population size of the "CLP-plant" subgroup as seen in \cref{tab:nordwindplakssstat}, recordings from $n=124$ speakers were chosen, aiming for an average age close to $9.30 \pm 2.60$. These speakers were similarly divided, with $20\%$ for testing and $80\%$ for training, and this procedure was repeated $20$ times.

\subsubsection*{Effect of Pathological Diversity}

The study further investigated the influence of pathology diversity on speaker verification performance. Consistent with data in \cref{fig:plakss_results}, the same number of speakers for both training and testing was maintained, with a focus on closely matching age distribution. By pooling all patient data, the study contrasted the results against a control group. As indicated in \cref{tab:nordwindplakssstat}, for children, both age and size consistency were achievable due to the extensive recordings from healthy subjects. Following the established protocol, $20$ iterations were conducted where $n=400$ speakers with a mean age of $10.29 \pm 0.13\%$ and a mean total duration of $26.55 \pm 0.58\%$ were selected for training. Meanwhile, 100 speakers with a mean age of $10.05 \pm 0.48\%$ and a mean total duration of $6.80 \pm 0.30\%$ were designated for testing from the combined "CLP-dnt" and "CLP-plant" patient groups. Concurrently, 400 speakers with a mean age of $11.72 \pm 0.10\%$ and a mean total duration of $24.08 \pm 0.55\%$ for training and $n=100$ speakers with a mean age of $11.70 \pm 0.33\%$ and a mean total duration of $6.03 \pm 0.32\%$ for testing were chosen from the "ctrl-plant-C" group.

\subsubsection*{Training Size’s Influence}

This section explored the effect of training set size on ASV system performance. Using recordings from different patient groups alongside a control set, the selection was determined by age and recording duration. To specifically assess training size impact, all $n=3,849$ available pathological and healthy speakers were amalgamated. Different quantities of speakers were randomly chosen for the routine training and evaluation steps: $n=50$, $500$, $1,500$, and $3,000$ speakers. For each group, $20\%$ was allocated to the test set and $80\%$ to the training set. Each sampling and evaluation cycle was reiterated $20$ times to consider random variations.

\subsubsection*{Intelligibility’s Effect}

The final phase was a correlation analysis, aiming to discern the relationship between speaker clarity (measured by intelligibility metrics) and ASV system performance metrics. This correlation explored the connection between EER results and WRR values throughout all experimental stages, offering insights into pathological speech nuances within speaker verification systems.

\subsection*{DL-Based ASV System}
\label{sec:methods}


Although DL-based methods, generally, outperform the classical speaker recognition methods, for instance, the i-vector approach~\cite{inproceedingsdeha, 5545402, 4960564}, in the context of text-independent speaker verification (TISV), the i-vector framework and its variants are still the state-of-the-art in some of the tasks~\cite{6853887, 10.1109/TASL.2010.2064307, inproceedingsasd, nist-2012, nist-2016}.
However, i-vector systems showed performance degradation when short utterances are met in enrollment/evaluation phase~\cite{inproceedingsasd}. Given that the children subset of our corpus contains a large amount of utterances with short lengths (less than $4$s), due to the nature of the PLAKSS test it makes sense for us to select a generalized TISV model, which can address our problem better. 
According to the results reported in~\cite{inproceedingsasd, 7846260, 7953194}, end-to-end DL systems achieved better performance compared to the baseline i-vector system~\cite{5545402}, especially for short utterances.
A major drawback of these systems is the time and cost required for training. Because of the nature of this study, we aimed at performing a considerable number of different experiments. Therefore, having a state-of-the-art end-to-end TISV model, which requires less training time is crucial. Thus, we chose to utilize the Generalized End-to-End (GE2E) TISV model proposed by Wan et al.~\cite{wan2017generalized}, which enabled us to process a large number of utterances at once and greatly decreased the total training and convergence time \cite{arasteh2020generalized}. 
The final embedding vector (d-vector) $\mathbf{e}_{ji}$ was the $L_2$ normalization of the network output and represents the embedding vector of the $j$th speaker’s $i$th utterance. The centroid of the embedding vectors from the $j$th speaker
$[\mathbf{e}_{j1}, ... , \mathbf{e}_{jM}]$ $\mathbf{c}_{j}$ was defined as the arithmetic mean of the embedding vectors of the $j$th speaker. 

The similarity matrix $\mathbf{S}_{ji,k}$ was defined as the scaled cosine similarities between each embedding vector $\mathbf{e}_{ji}$ to all centroids $\mathbf{c}_{k}$ $(1 \leq j,k \leq N$, and $1 \leq i \leq M)$. Furthermore, removing $\mathbf{e}_{ji}$ when computing the centroid of the true speaker made training stable and helps avoid trivial solutions~\cite{wan2017generalized}. Thus, the similarity matrix could be written as following:

\begin{equation}
  \mathbf{S}_{ji,k} =
    \begin{cases}
      w \cdot \cos(\mathbf{e}_{ji}, \mathbf{c}_{j}^{(-i)})+b & \text{if $k=j$}\\
       w \cdot \cos(\mathbf{e}_{ji}, \mathbf{c}_{k})+b & \text{otherwise},
    \end{cases}       
\end{equation}
with $w$ and $b$ being the trainable weights and biases. 
As we can see, unlike most of the end-to-end methods, rather than a scalar value, GE2E builds a similarity matrix that defines the similarities between each $\mathbf{e}_{ji}$ and all centroids $\mathbf{c}_{k}$.

We put a SoftMax on $\mathbf{S}_{ji,k}$ for $k = 1, ... , N$ that makes the output equal to one if $k = j$, otherwise makes the output equal to zero. Thus, the loss on each embedding vector $\mathbf{e}_{ji}$ could be defined as:

\begin{equation}
    L(\mathbf{e}_{ji})= -\mathbf{S}_{ji,j} + \log\sum_{k=1}^N\exp(\mathbf{S}_{ji,k}).
\label{eq:loss_single}
\end{equation}
Finally, the GE2E loss $L_G$ is the mean of all losses over the similarity matrix ($1 \leq j \leq N$, and $1 \leq i \leq M$):
    	\begin{equation}
   L_G=\frac{1}{M\cdot N}\sum_{j,i}L(\mathbf{e}_{ji}).
\end{equation}

\subsection*{Training Steps}

By specifying a set of clear training and evaluation steps for all the experiments, we aimed at standardizing our experiments and preventing influences of non-pathology factors.
We followed a similar data pre-processing scheme as in \cite{arasteh2020generalized, wan2017generalized, 7178863} and pruned the intervals with sound pressures below $30$ db. Afterward, we performed voice activity detection~\cite{Ramirez07} to remove the silent parts of the utterances, with a window length of $30$ ms, a maximum silence length of $6$ ms, and a moving average window of the length $8$ ms. Removing silent parts, we ended up with partial utterances of each utterance, where we merely chose the partial utterances which have a minimum length of $1,825$ ms for training, due to the fact that our dataset contained utterances with a $16$ kHz sampling rate. Our final feature representations were 40-dimensional log-Mel-filterbank energies, where we used a window length of $25$ ms with steps of $10$ ms and, i.e., a short time Fourier transform (STFT) of size of 512.
To prepare training data batches, similar to~\cite{wan2017generalized}, we selected $N$ different speakers and fetched $M$ different utterances for every selected speaker to create a training batch. Furthermore, while we could have batches with different partial utterance lengths, due to the fact that all the partial utterances of each training batch should have the same length \cite{wan2017generalized}, we randomly segmented all the partial utterances of each training batch to have the same length. All the procedures to pre-process raw input waveforms and prepare training batches are explained in \cref{alg:train}.

\begin{figure}[t]
    \centering
        \includegraphics[width=\linewidth, height=2.5cm, keepaspectratio]{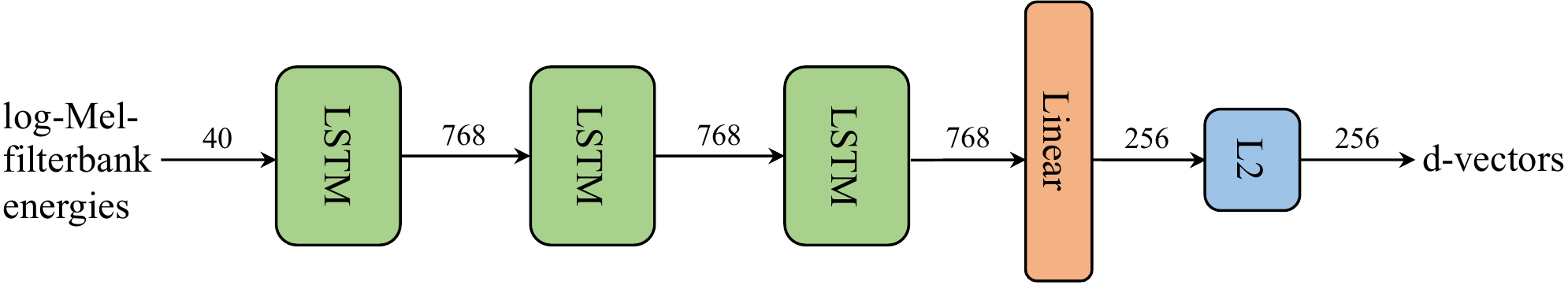}
    \caption{The architecture of the utilized text-independent speaker verification model. The inputs of the network are 40-dimensional log-Mel-filterbank energies, which are the results of performing data pre-processing steps on raw utterances. The numbers above each arrow represent the feature dimensions at each step. The final 256-dimensional d-vectors are the $L_2$ normalization of the network outputs.}
    \label{fig:arch}
\end{figure}

Our network architecture, which is shown in \cref{fig:arch}, consisted of 3 long short-term memory (LSTM) layers~\cite{10.1162/neco.1997.9.8.1735} with 768 hidden nodes followed by a linear projection layer in order to get to the 256-dimensional embedding vectors~\cite{articlesak}.
The $L_2$ norm of gradient was clipped at 3~\cite{Pascanu2012UnderstandingTE}. 
In order to prevent coincidental training cases, the Xavier normal initialization~\cite{articlexavier} was applied to the network weights and the biases were initialized with zeros for all the experiments.
The Adam~\cite{kingma2014adam} optimizer was selected to optimize the model. Depending on each individual experiment and more specifically, its training set, we chose a different learning rate per experiment from $10^{-5}$ to $10^{-4}$, in a way that the network converges the best. 
For all of the experiments, during training, we selected $N=16$ speakers and $M=4$ partial utterances per speaker.
Moreover, no pre-trained model was used during training of each experiment, and we always started training from scratch with the same initialization.

\begin{algorithm}
 \For{all training batches}{
  $-$ randomly choose an integer $L$ within $[140, 180]$\;
 \For{all train speakers}{
 $-$ randomly choose $N$ speakers\;
 \For{all $N$ speakers}{
   $-$ initialize empty set S\;
  \For{all utterances}{
 $-$ normalize the volume\;
 $-$ perform VAD with max\_silence\_length $=6$ ms and window\_length$ =30$ ms\;
 $-$ prune the intervals with sound pressures below $30$ db\;
 \For{all resulting partial utterances}{
   \uIf{partial utterance's length $>$180 frames}{
    $-$ add partial utterance to $S$\;}}}
  $-$ randomly select $M$ partial utterances from $S$\;
   \For{all selected partial utterances}{
    $-$ perform STFT on the partial utterance\;
    $-$ take magnitude squared of result\;
    $-$ transform to the Mel scale\;
    $-$ take the logarithm\;
    $-$ randomly segment an interval with $L$ frames\;}

  }}}
\caption{Training data preparation steps.}
\label{alg:train}
\end{algorithm}

\subsection*{Evaluation Method}

For the evaluation of the trained networks, we followed the same data pre-processing steps as for training, with the only difference that, during evaluation, we concatenated all the partial utterances corresponding to each utterance before feeding them to the network. Then, as proposed by Wan et al. \cite{wan2017generalized}, we applied a sliding window of a fixed size ($160$ frames) with 50\% overlap to the concatenated utterances and performed an element-wise averaging on the d-vectors to get the final d-vector representation of the test utterance.
Furthermore, Tayebi Arasteh et al.\cite{arasteh2020generalized} showed that the choice of the parameter $M$ for evaluation is an influencing factor in the resulting prediction, i.e., the more enrollment utterances, usually, the better prediction for test utterances. 
Therefore, we decided to report the results for $M=2$, where we have only one enrollment utterance (during the calculation of centroid of the true speaker, we excluded the utterance itself as proposed by Wan et al. \cite{wan2017generalized}), as we did not see large deviations for other choices of $M$ in results which cannot be reported here for brevity. The results for $M=4$ are reported in the supplementary information (see Table S1).
For each experiment, we chose the batch size $N$ to be equal to the total number of the test speakers during evaluation.
To prevent the effect of random sampling in choosing recordings of training and testing for different experiments, we repeated each experiment 20 times and calculated the statistics accordingly. All the steps to pre-process raw input waveforms for enrollment and evaluation as well as the steps for preparing final d-vectors are stated in \cref{alg:test}.

\begin{algorithm}
 \For{all enrollment and evaluation speakers}{
  \For{all utterances}{
     $-$ initialize empty set $A$\;
   $-$ normalize the volume\;
 $-$ perform VAD with max\_silence\_length $=6$ ms and window\_length $=30$ ms\;
 $-$ prune the intervals with sound pressures below $30$ db\;
 \For{all resulting partial utterances}{
   \uIf{partial utterance's length $>$ 180 frames}{
   $-$ add the partial utterance to $A$\;
   }}
  $-$ concatenate the elements of $A$\;
  $-$ perform STFT on the concatenated utterance\;
  $-$ take the magnitude squared of the result\;
  $-$ transform to the Mel scale\;
  $-$ take the logarithm\;

   $-$ set $t=0$\;
   $-$ initialize empty set $D$\;
   \While {$t+160 <$ length of the utterance}{

    $-$ select the interval within $[t, t+160]$ frames of the utterance\;
    $-$ feed the selected utterance to the trained network to obtain the corresponding d-vector\;
    $-$ $L_2$-normalize the d-vector\;
    $-$ add the normalized d-vector to $D$\;
    $-$ $t=t+80$\;
   }
   $-$ perform element-wise average of elements of $D$ to obtain the final utterance d-vector\;
   }}
\caption{Enrollment and evaluation data preparation followed by d-vector creation steps.}
\label{alg:test}
\end{algorithm}

\subsection*{Quantitative Analysis Metric}

As our main quantitative evaluation metric, we chose EER, which is used to predetermine the threshold values for its false acceptance rate (FAR) and its false rejection rate (FRR)~\cite{vanLeeuwen2007, Hansen2015SpeakerRB}. It looks for a threshold for similarity scores where the proportion of genuine utterances which are classified as imposter, i.e., the FRR is equal to the proportion of imposters classified as genuine, i.e., the FAR \cite{arasteh2020generalized}.
The similarity metric, which we use here, is the cosine distance score, which is the normalized dot product of the speaker model and the test d-vector:

\begin{equation}
    \cos(\mathbf{e}_{ji} , \mathbf{c}_{k})=\frac{\mathbf{e}_{ji} \cdot \mathbf{c}_{k}}{\lVert \mathbf{e}_{ji} \rVert \cdot \lVert \mathbf{c}_{k} \rVert}.
\label{eq:cosine}
\end{equation}
The higher the similarity score between $\mathbf{e}_{ji}$ and $\mathbf{c}_{k}$ is, the more similar they are. We report the EER values in percent throughout this paper.

\subsection*{Statistical Analysis}

Descriptive statistics are reported as median and range, or mean $\pm$ standard deviation, as appropriate. Normality was tested using Shapiro-Wilk test \cite{SHAPIRO1965}. A two-tailed unpaired t-test was used to compare two groups of EER data with Gaussian distributions. A P$ \leqslant 0.05$ was considered statistically significant.


\subsection*{Code Availability}

The full source code including training and evaluation of the recurrent neural networks, data pre-processing and feature extraction steps, and analysis of the results are publicly available at \url{https://github.com/tayebiarasteh/pathology_ASV}. All the code is developed in Python 3.9. The PyTorch 1.13 framework is used for deep learning.

\subsection*{Data Availability}

The speech dataset used in this study is not publicly available as it is internal data of patients of the University Hospital Erlangen. A reasonable request to the corresponding author is required for accessing the data on-site at the University Hospital Erlangen in Erlangen, Bavaria, Germany.

\subsection*{Hardware}

The hardware used in our experiments were Intel CPUs with 18 and 32 cores and 32 GB RAM and Nvidia GPUs of GeForce GTX 1080 Ti, V100, RTX 6000, Quadro 5000, and Quadro 6000 with 11 GB, 16 GB,  24 GB, 32 GB, and 32 GB memories, respectively.


\bibliography{sample}


\section*{Funding}

This study was funded by Friedrich-Alexander-University Erlangen-Nuremberg, Medical Valley e.V., and Siemens Healthineers AG within the framework of d.hip campus.

\section*{Author Contributions}

S.T.A. cleaned, organized, and pre-processed the data, developed the software, conducted the experiments, performed the statistical analysis, and wrote the manuscript. T.W. prepared the data and contributed to the editing. M.S. guided the data collection, counseled on clinical relevance, and contributed to the editing. E.N. guided the study design and contributed to the editing. A.M. supported the conception of the study and the experiments, corrected, and edited the manuscript. S.H.Y. designed the study and greatly guided the writing, corrected, and edited the manuscript.


\section*{Competing Interests}

The authors declare no competing interests.

\beginsupplement
\section*{Supplementary Information}

\begin{table}[ht]
\centering
\begin{tabular}{>{\arraybackslash}m{3.0cm}>{\centering\arraybackslash}m{1.5cm}>{\centering\arraybackslash}m{2.1cm}>{\centering\arraybackslash}m{2.1cm}>{\centering\arraybackslash}m{2.1cm}>{\centering\arraybackslash}m{2.1cm}>{\centering\arraybackslash}m{2.1cm}}
\toprule
\textbf{} & \textbf{Total num speakers} & \textbf{EER [\%] $M=4$} & \textbf{Total duration train [hours]} & \textbf{Total duration test [hours]} & \textbf{Age train [years]} & \textbf{Age test [years]} \\
\midrule
\textbf{Adults} & &  &    &  &  &  \\
dysglossia-dnt-85 & $85$ & $2.17\pm 0.52$ & $3.16\pm 0.34$  & $0.75\pm 0.20$ & $58.58\pm 14.97$ & $57.84\pm 15.36$\\
dysarthria-plant-85 & $85$ & $1.86\pm 0.81$ & $1.71\pm 0.15$  & $0.45\pm 0.08$ & $60.12\pm 15.58$ & $60.29\pm 14.60$ \\
dysphonia-logi-85 & $85$ & $1.32\pm 0.59$ &  $1.32\pm 0.05$ & $0.34\pm 0.05$ & $58.91\pm 11.31$ & $59.49\pm 9.85$\\
ctrl-plant-A-85 & $85$ & $2.04\pm 0.67$ & $1.28\pm 0.03$ & $0.32\pm 0.03$ & $23.78\pm 15.47$ & $23.98\pm 14.33$\\
\midrule
\midrule
\textbf{Children} & &  &    &   &  \\
CLP-dnt-124 & $124$ & $3.67\pm 0.61$ & $6.79\pm 0.44$  & $1.77\pm 0.30$ & $9.48\pm 3.20$ & $9.31\pm 3.06$\\
CLP-plant-124 & $124$ & $5.01\pm 0.69$ & $6.31\pm 0.10$  & $1.57\pm 0.10$ & $9.27\pm 2.54$  & $9.20\pm 2.40$\\
ctrl-plant-C-124 & $124$ & $3.98\pm 0.86$ &  $5.90\pm 0.22$ & $1.44\pm 0.15$ & $11.14\pm 3.21$ & $11.11\pm 3.08$ \\
all-children-124 & $124$ & $3.14\pm 0.66$ &  $6.03\pm 0.29$ & $1.54\pm 0.12$ & $10.54\pm 3.17$ & $10.49\pm 3.08$ \\
CLP-dnt-plant-500 & $500$ & $1.88\pm 0.21$ & $26.55\pm 0.58$  & $6.80\pm 0.30$ & $10.29\pm 4.73$ & $10.05\pm 4.43$\\
ctrl-plant-C-500 & $500$ & $1.93\pm 0.15$ &  $24.08\pm 0.55$ & $6.03\pm 0.32$ & $11.72\pm 3.69$ & $11.70\pm 3.65$ \\
\midrule
\midrule
\textbf{Logarithmic effect} & &  &    &   &  \\
all-spk-50 & $50$ & $3.75\pm 1.18$ & $1.98\pm 0.21$  & $0.56\pm 0.10$ & $25.50\pm 22.70$ & $22.06\pm 18.27$\\
all-spk-500 & $500$ & $1.21\pm 0.12$ & $20.22\pm 0.73$  & $4.99\pm 0.34$ & $24.87\pm 22.86$  & $24.60\pm 22.48$\\
all-spk-1500 & $1,500$ & $0.71\pm 0.07$ &  $60.58\pm 1.17$ & $15.03\pm 0.59$ & $25.22\pm 23.12$ & $25.17\pm 23.13$ \\
all-spk-3000 & $3,000$ & $0.55\pm 0.03$ & $120.86\pm 1.15$ & $30.34\pm 0.85$ & $25.08\pm 23.06$ & $25.16\pm 23.07$  \\
\bottomrule
\end{tabular}
\caption{Statistics of the training and test sets of all the experiments. In all the experiments, 20\% of the speakers were assigned to test and 80\% of the speakers to training. Abbreviations: dysglossia: Patients with dysglossia who underwent prior maxillofacial surgery; dysarthria: Patients diagnosed with dysarthria; dysphonia: Patients with voice disorders; CLP: Children with cleft lip and palate; dnt: Recordings from the "dnt Call 4U Comfort" headset; plant: Recordings via Plantronics Inc. headset; logi: Recordings via Logitech International S.A. headset; ctrl: Control group. The labels "-A" and "-C" respectively indicate adult and children subsets. Numbers appended, such as "-85" in "dysglossia-dnt-85", represent the total speaker count for that experiment. "all-spk" designates experiments combining all dataset speech signals from both adults and children, and both pathological and healthy subjects.}
\label{tab:chosennordwindplakssstat}
\end{table}

\end{document}